\documentclass[fleqn,usenatbib]{mnras}

\usepackage[T1]{fontenc}
\usepackage{ae,aecompl}

\usepackage{graphicx}	
\usepackage{amsmath}	
\usepackage{amssymb}	

\title[The Reactivation of MBC 324P/La Sagra]{The Reactivation of Main-Belt Comet 324P/La Sagra (P/2010 R2)}

\author[H.~H.\ Hsieh and S.~S.\ Sheppard]{
Henry H.\ Hsieh,$^{1,2}$\thanks{E-mail: hhsieh@asiaa.sinica.edu.tw}
Scott S.\ Sheppard$^{3}$
\\
        $^1$Institute of Astronomy and Astrophysics, Academia Sinica, P.O.\ Box 23-141, Taipei 10617, Taiwan\\
        $^2$Planetary Science Institute, 1700 East Fort Lowell Rd., Suite 106, Tucson, AZ 85719, USA\\
        $^3$Carnegie Institution for Science, 5241 Broad Branch Rd.\ NW, Washington, DC 20015
}

\date{Accepted 2015 August 24. Received 2015 August 24; in original form 2015 August 09}

\pubyear{2015}

\begin{document}
\label{firstpage}
\pagerange{\pageref{firstpage}--\pageref{lastpage}}
\maketitle

\begin{abstract}
We present observations using the Baade Magellan and Canada-France-Hawaii telescopes showing that main-belt comet 324P/La Sagra, formerly known as P/2010 R2, has become active again for the first time since originally observed to be active in 2010-2011.  The object appears point-source-like in March and April 2015 as it approached perihelion (true anomaly of $\nu\sim300^{\circ}$), but was $\sim$1~mag brighter than expected if inactive, suggesting the presence of unresolved dust emission.  Activity was confirmed by observations of a cometary dust tail in May and June 2015.  We find an apparent net dust production rate of ${\dot M_d}\lesssim0.1$~kg~s$^{-1}$ during these observations.  324P is now the fourth main-belt comet confirmed to be recurrently active, a strong indication that its activity is driven by sublimation.  It now has the largest confirmed active range of all likely main-belt comets, and also the most distant confirmed inbound activation point at $R$$\,\sim\,$2.8~AU.  Further observations during the current active period will allow direct comparisons of activity strength with 324P's 2010 activity.
\end{abstract}




\section{Introduction}
\subsection{Background}
\label{section:background}

Active asteroids are rare objects (fewer than two dozen are currently known) that exhibit comet-like dust emission, yet are located inside the orbit of Jupiter and have asteroid-like orbits, typically defined as having a Tisserand parameter of $T_J\gtrsim3$ \citep{jew15b}.  They consist of main-belt comets \citep[MBCs;][]{hsi06}, whose activity is believed to be driven by volatile ice sublimation, and disrupted asteroids, whose mass loss is believed instead to be produced by an impact, rotational destabilization, or other physical mechanisms \citep{hsi12a,jew15b}.

Distinguishing between MBCs and disrupted asteroids is non-trivial as no spectroscopic detection of outgassing in active asteroids has been made to date \citep[e.g.,][]{jew09,bod11,hsi12a,hsi12b,hsi12c,hsi13,lic11,lic13,dev12,oro13}, likely due to low gas production rates that are undetectable by current Earth-based facilities.  Water vapor outgassing has been detected for the main-belt dwarf planet (1) Ceres \citep{kup14}, but visible dust emission has never been seen, and with a diameter of 950~km, Ceres is also far larger than any of the known (km-scale) MBCs.

The lack of direct detections of outgassing for most MBCs means that action of sublimation must be inferred indirectly.  Steady or increasing long-lived activity is suggestive, but is not indisputable, evidence of sublimation-driven activity, while impulsive dust emission and rapid fading is suggestive of an  impact event \citep[e.g.,][]{jew11}.  Numerical dust modeling can be used to infer the presence of long-lasting activity, strongly suggesting the action of sublimation \citep[e.g.,][]{hsi04,hsi12b,jew14a,jew14b,mor11,mor13}, but such modeling is susceptible to parameter degeneracies. Complicating matters, some disruptive effects like spin-up disruption could, in principle, also produce long-lasting dust emission.

The property that is considered the most definitive, if still indirect, evidence of sublimation-driven activity is recurrent activity near perihelion separated by intervening periods of inactivity \citep{hsi12a,jew15b}.  Such behavior cannot be easily explained by impacts, rotational disruption, or any other mechanism suspected of driving disrupted asteroid activity.  Prior to the work presented here, recurrent activity has been observed for just three MBCs --- 133P/Elst-Pizarro, 238P/Read, and 313P/Gibbs \citep{hsi04,hsi11,hsi15b,jew15}.

\begin{figure*}
\label{figure:fig_images_orbplot}
\includegraphics[width=6.5in]{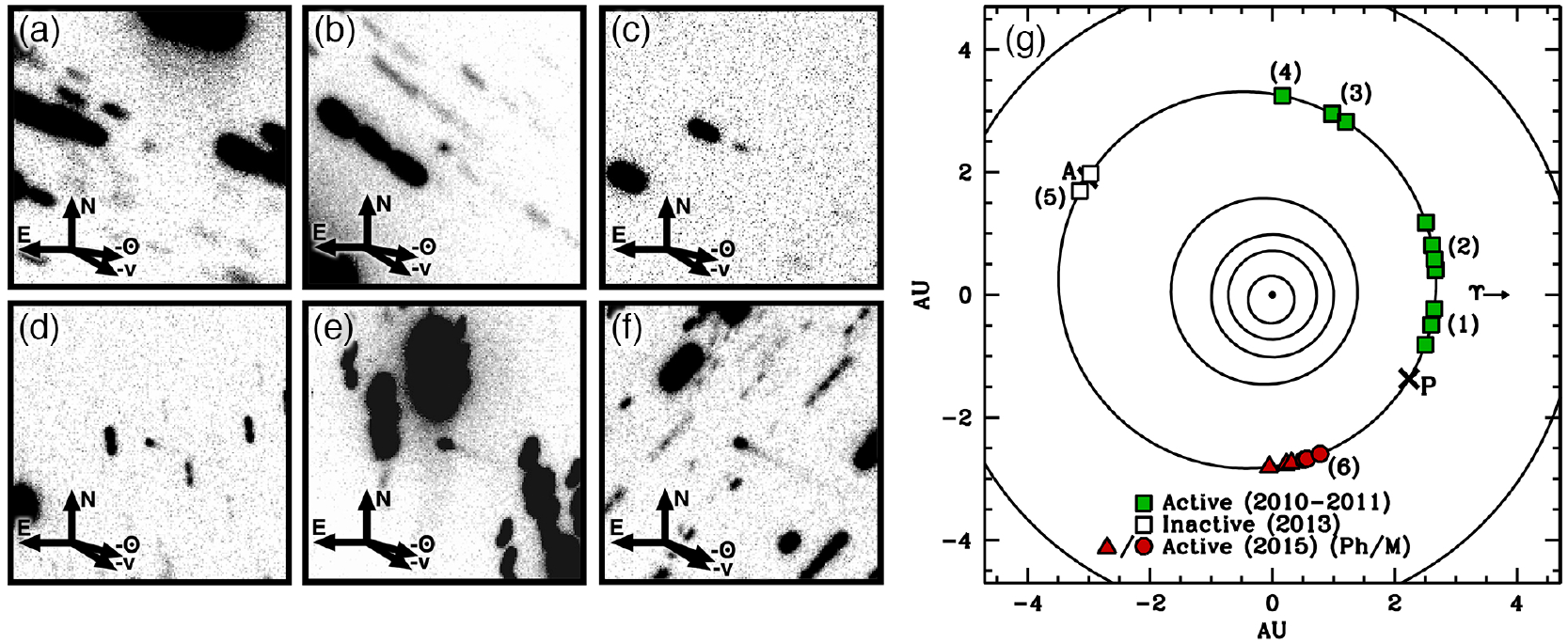}
\caption{Composite images constructed from observations obtained on (a) 2015 March 21, (b) 2015 April 18, (c) 2015 April 26, (d) 2015 May 21, (e) 2015 May 22, and (f) 2015 June 13 (Table~\ref{table:obslog}).  The object is at the center of each $30''\times30''$ panel, and north (N), east (E), the antisolar direction ($-\odot$), and the negative heliocentric velocity vector ($-v$), as projected in the plane of the sky, are marked with arrows. (g) Orbital plot showing positions of 324P with the Sun (black dot) at the center of the plot, and the orbits of Mercury, Venus, Earth, Mars, 324P, and Jupiter shown as black lines.  Perihelion (P) and aphelion (A) are marked with crosses.  Green squares mark the positions of observations where 324P was visibly active in 2010 and 2011 \citep{hsi12c,hsi14}, open squares mark where 324P appeared inactive in 2013 \citep{hsi14}, and red symbols mark where 324P was observed to be active again in 2015 (this work), where triangles indicate where activity was photometrically detected and circles indicate where activity was morphologically detected.
}
\end{figure*}

\subsection{324P/La Sagra\label{section:324p}}

Comet 324P/La Sagra was discovered as P/2010 R2 on 2010 September 14 \citep{nom10}.  It has a semimajor axis of $a$$\,=\,$3.099~AU, an eccentricity of $e$$\,=\,$0.154, an inclination of $i$$\,=\,$21.396$^{\circ}$, and a Tisserand parameter of $T_J$$\,=\,$3.099, dynamically placing it in the main asteroid belt.  Numerical modeling indicated that the observed emission could be simulated by anisotropic mass loss of $\sim$3$-$4~kg~s$^{-1}$ over $>$7 months following perihelion, with an emission source near the object's south pole and an obliquity of $\sim$90$^{\circ}$ \citep{mor11}.

Precovery and follow-up data showed that 324P brightened by $>$1~mag between 2010 August and December, further implying that dust production was ongoing over that period \citep{hsi12c}.  December observations also revealed the appearance of both an antisolar dust tail and a dust trail aligned with the object's orbit plane.  This morphology is strongly diagnostic of a prolonged emission event because it indicates the simultaneous presence of large, slow-moving particles emitted a long time in the past (the dust trail), and small, fast-dissipating particles that must have been emitted very recently (the antisolar tail) \citep{hsi12b}.

The evidence for long-lived dust emission strongly implies that the comet-like mass loss observed for 324P in 2010 was driven by the sublimation of ice, and was not caused by an impulsive mechanism such as an impact.  In this paper, we present significant additional support for this conclusion in the form of observations of 324P's reactivation, making it now the fourth MBC confirmed to exhibit recurrent activity.

\section{Observations\label{observations}}

We observed 324P between 2015 March and 2015 June with the 6.5~m Baade Magellan telescope at Las Campanas, and the 3.54~m Canada-France-Hawaii Telescope (CFHT) on Mauna Kea (Table~\ref{table:obslog}).  We employed the Inamori Magellan Areal Camera and Spectrograph (pixel scale of $0\farcs200$~pixel$^{-1}$) for Baade observations \citep{dre11}, and MegaCam ($0\farcs187$~pixel$^{-1}$) on CFHT \citep{bou03}.  Non-sidereal tracking was used for all observations.  A broadband filter spanning the standard $V$- and $R$-band wavelength ranges was used for all Magellan observations, while in April, a SDSS-like $r'$-band filter was also used.  Bias subtraction and flat-field reduction (using dithered images of the twilight sky) were performed for Magellan data, while photometric calibration was performed using SDSS standard stars with colors similar to moderately red asteroids.  Photometry of $VR$- and $r'$-band April Magellan data produced consistent results, providing assurance that our other $VR$-band photometry is reliable.  All CFHT observations utilized a $r'$-band filter. Reduction of CFHT data was performed by the Elixir pipeline \citep{mag04}, while calibration was performed using field stars from the Pan-STARRS1 catalog \citep{sch12,ton12,mag13} as photometric references.  Equivalent mean apparent $R$-band magnitudes measured for 324P using $2\farcs0$ apertures for all data are listed in Table~\ref{table:obslog}.

324P appears stellar in March and April observations (Figure~\ref{figure:fig_images_orbplot}; seeing of $\theta_s\sim0\farcs8-1\farcs0$) where we measure point spread function (PSF) widths for the object (measured perpendicularly to the direction of the non-sidereal tracking) that are similar to those of nearby field stars.  Full PSF profiles for these data are too noisy to provide any useful assessment of whether low-level activity is present. The object appears visibly cometary in May and June observations, with PSF widths $\sim$10\% larger than those of nearby field stars ($\theta_s\sim0\farcs6-0\farcs9$) and exhibiting a tapered tail extending as far as $\sim$15$''$ in the antisolar direction.

\setlength{\tabcolsep}{3.5pt}
\begin{table*}
	\centering
	\caption{Observations}
	\label{table:obslog}
	\begin{tabular}{lcrrcrrrrccccc} 
		\hline
		{\bf UT Date} & {\bf Tel.$^{a}$}
 & {\bf N$^{b}$}
 & {\bf t$^{c}$}
 & {\bf Filter}
 & {\bf $\nu$$^{d}$}
 & {\bf $R$$^{e}$}
 & {\bf $\Delta$$^{f}$}
 & {\bf $\alpha$$^{g}$}
 & {\bf $m_{R,n}$$^{h}$}
 & {\bf $m_{R,t}$$^{i}$}
 & {\bf $H_{R,t}$$^{j}$}
 & {\bf $Af\rho$$^{k}$}
 & {\bf $M_d$$^{l}$} \\
\hline
2010 Jun 26     & \multicolumn{4}{l}{\it Perihelion.......................} &   0.0 & 2.623 & 2.228 & 22.3 & --- & --- & --- & --- & --- \\ 
2015 Mar 21     & Magellan & 2 &  550 & $VR$ & 299.6 & 2.809 & 2.961 & 19.7 & 23.2$\pm$0.2 & ---          & 17.6$\pm$0.2 & $2.7\pm0.1$ & $(3.1\pm1.1)\times10^6$ \\ 
2015 Apr 18     & Magellan & 1 &  400 & $VR$ & 305.8 & 2.773 & 2.553 & 21.2 & 22.8$\pm$0.1 & ---          & 17.5$\pm$0.1 & $3.5\pm0.2$ & $(3.7\pm0.6)\times10^6$ \\ 
2015 Apr 18     & Magellan & 2 &  800 & $r'$ & 305.8 & 2.773 & 2.553 & 21.2 & ...          & ---          & ...          & ...         & ...                     \\ 
2015 Apr 26     & CFHT     & 1 &  180 & $r'$ & 307.7 & 2.763 & 2.434 & 21.1 & 22.7$\pm$0.2 & ---          & 17.6$\pm$0.2 & $3.6\pm0.2$ & $(3.6\pm1.2)\times10^6$ \\ 
2015 May 16     & CFHT     & 4 &  720 & $r'$ & 312.2 & 2.740 & 2.158 & 19.6 & 22.5$\pm$0.1 & ---          & 17.7$\pm$0.1 & $3.6\pm0.2$ & $(2.9\pm0.6)\times10^6$ \\ 
2015 May 17     & CFHT     & 3 &  540 & $r'$ & 312.4 & 2.739 & 2.145 & 19.5 & 22.6$\pm$0.1 & ---          & 17.8$\pm$0.1 & $3.3\pm0.2$ & $(2.3\pm0.5)\times10^6$ \\ 
2015 May 21     & CFHT     & 3 &  540 & $r'$ & 313.3 & 2.734 & 2.094 & 18.9 & 22.3$\pm$0.1 & ---          & 17.6$\pm$0.1 & $4.1\pm0.2$ & $(3.5\pm0.6)\times10^6$ \\ 
2015 May 22     & Magellan & 3 & 1050 & $VR$ & 313.5 & 2.733 & 2.084 & 18.7 & 22.1$\pm$0.2 & ---          & 17.4$\pm$0.2 & $4.9\pm0.2$ & $(4.7\pm1.4)\times10^6$ \\ 
2015 May 23     & CFHT     & 3 &  540 & $r'$ & 313.8 & 2.732 & 2.069 & 18.5 & 22.4$\pm$0.1 & ---          & 17.7$\pm$0.1 & $3.7\pm0.2$ & $(2.7\pm0.5)\times10^6$ \\ 
2015 Jun 10     & CFHT     & 3 &  540 & $r'$ & 317.9 & 2.713 & 1.873 & 14.5 & 22.0$\pm$0.1 & 21.9$\pm$0.1 & 17.6$\pm$0.1 & $4.2\pm0.2$ & $(3.5\pm0.6)\times10^6$ \\ 
2015 Jun 13     & CFHT     & 7 & 1260 & $r'$ & 318.6 & 2.710 & 1.846 & 13.7 & 22.1$\pm$0.1 & 21.8$\pm$0.1 & 17.5$\pm$0.1 & $3.7\pm0.2$ & $(3.7\pm0.6)\times10^6$ \\ 
2015 Nov 30  & \multicolumn{4}{l}{\it Perihelion.......................} &   0.0 & 2.620 & 2.946 & 19.3 & --- & --- & --- & --- & --- \\ 
		\hline
\multicolumn{14}{l}{$^a$~Telescope}\\
\multicolumn{14}{l}{$^b$~Number of exposures.}\\
\multicolumn{14}{l}{$^c$~Total integration time, in s.}\\
\multicolumn{14}{l}{$^d$~True anomaly, in degrees.}\\
\multicolumn{14}{l}{$^e$~Heliocentric distance, in AU.}\\
\multicolumn{14}{l}{$^f$~Geocentric distance, in AU.}\\
\multicolumn{14}{l}{$^g$~Solar phase angle (Sun-object-Earth), in degrees.}\\
\multicolumn{14}{l}{$^h$~Equivalent mean apparent $R$-band nucleus magnitude, measured with a $2\farcs0$ aperture.}\\
\multicolumn{14}{l}{$^i$~Equivalent mean apparent $R$-band total magnitude, measured using apertures encompassing the entire coma and tail, if applicable.}\\
\multicolumn{14}{l}{$^j$~Absolute $R$-band total magnitude (at $R=\Delta=1$~AU and $\alpha=0\degr$), assuming IAU $H,G$ phase-darkening where $G=0.17$.}\\
\multicolumn{14}{l}{$^k$~$Af\rho$ values (computed using $2\farcs0$ photometry apertures), in cm.}\\
\multicolumn{14}{l}{$^l$~Estimated total dust mass, in kg, assuming $\rho_{d}\sim2500$~kg~m$^3$.}\\
	\end{tabular}
\end{table*}


\section{Results and Analysis}
\label{section:results}

\subsection{Photometric Analysis}
\label{section:phot_results}

We compute absolute $R$-band magnitudes for 324P for each set of observations (i.e., at $R$$\,=\,$$\Delta$$\,=\,$1~AU and $\alpha$$\,=\,$$0^{\circ}$) (Table~\ref{table:obslog}), assuming inverse-square law fading and an $H,G$ phase function with $G$$\,=\,$0.17 as computed for 324P's nucleus \citep{hsi14}. Emitted dust may not have the same photometric behavior as the nucleus though, making this assumption a source of uncertainty.  For reference, we also compute $Af\rho$ values \citep{ahe84}.
In all of our observations, 324P's observed absolute magnitude is $\sim$1~mag brighter than the absolute magnitude of the inactive nucleus \citep[$H=18.4$~mag;][]{hsi14}, suggesting the presence of unresolved dust emission even in data in which activity is not visually detected (i.e., in March and April).

\begin{figure*}
	\includegraphics[width=6.5in]{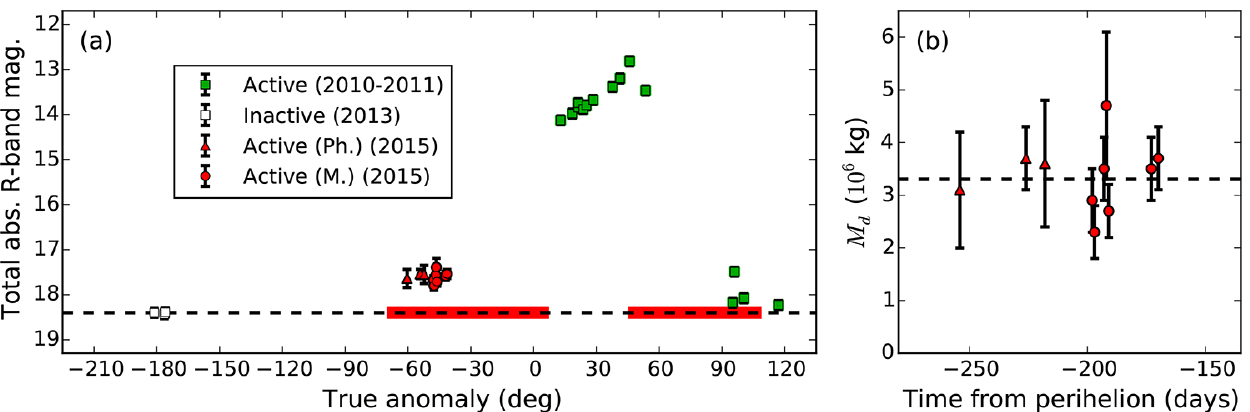}
    \caption{(a) Total absolute $R$-band magnitude of 324P during its 2010-2011 active period (green squares) and its 2015 active period to date (red triangles where activity was photometrically detected, and red circles where activity was morphologically detected) plotted as a function of true anomaly.  Also plotted are magnitudes measured in 2013 when 324P was apparently inactive (open squares).  The expected magnitude of the inactive nucleus is marked with a horizontal dashed black line.  Thick red lines indicate the true anomaly ranges during which the object will be observable from Earth (i.e., with a solar elongation of $>50^{\circ}$) during its current perihelion passage, where the first segment corresponds to 2015 March to 2015 December, and the second segment corresponds to 2016 June to 2017 March. (b) Total dust masses measured for 324P from 2015 March to 2015 June plotted as a function of time until perihelion.  A dashed line shows a linear fit to the data, reflecting the average net dust production rate over this period (see Section~\ref{section:prev_obs}).
}
    \label{figure:fig_photevolution_visibility}
\end{figure*}

\subsection{Comparison to Previous Observations\label{section:prev_obs}}

In Figure~\ref{figure:fig_photevolution_visibility}, we plot the total absolute $R$-band magnitudes of 324P (including flux from the entire coma and tail), $H_{R,tot}$, corrected for distance and phase angle effects, from observations from 2010 to 2015 \citep[][this work]{hsi12c,hsi14}.  From observed photometric excesses above the expected brightness of 324P's inactive nucleus, we estimate the total mass, $M_d$, of visible ejected dust using
\begin{equation}
M_d = {4\over 3}\pi r^2_N {\bar a}\rho_d \left({1-10^{0.4(H_{R,tot}-H_R)} \over 10^{0.4(H_{R,tot}-H_R)} }\right)
\end{equation}
where $r_N = 0.55$~km is the estimated effective nucleus radius for 324P \citep{hsi14}.  We assume a dust grain density of $\rho_d=2500$~kg~m$^{-3}$, consistent with CI and CM carbonaceous chondrites, which are associated with primitive C-type objects like the MBCs \citep{bri02}, and an effective mean dust grain radius of $\bar a$$\,=\,$1~mm \citep[assuming a power-law particle size distribution from $\mu$m- to cm-sized particles determined from dust modeling;][]{mor11,hsi14}.  Dust masses calculated in this way are shown in Table~\ref{table:obslog}.

We plot total dust masses calculated for 324P in our 2015 data in Figure~\ref{figure:fig_photevolution_visibility}b.  We also fit a linear function to these data to estimate the average net dust production rate, ${\dot M_d}$, over this period, finding ${\dot M_d}\lesssim0.1$~kg~s$^{-1}$ (subject to uncertainties due to the unknown dust phase function properties and rotational state of the nucleus at the time of our observations).
Using analogous calculations, \citet{hsi14} reported a dust production rate of ${\dot M_d}\sim30$~kg~s$^{-1}$ in 2010.
These net rates represent the difference between the true dust production rate and the dust dissipation rate, meaning that they represent lower limits to the true dust production rates.  Numerical dust modeling is needed to calculate more accurate dust production rates, but meaningful modeling will require additional observations (cf.\ Section~\ref{section:discussion}), placing it beyond the scope of this preliminary study.  Adopting these computed rates as order-of-magnitude estimates for now, we find that the net dust production rate measured in 2015 is at least $\sim$2 orders of magnitude smaller than that measured in 2010.

Over the 2010 observations considered here, 324P covered a 
heliocentric distance range of $R$$\,\sim\,$2.65$\,-\,$2.75$\,$AU, and 
$R$$\,\sim\,$2.70$\,-\,$2.80$\,$AU during our 2015 observations (covering similar phase angle ranges during each period).  Following \citet{hsi15a}, we find an equilibrium surface temperature (assuming grey-body water ice sublimation) of $T_{eq}$$\,\sim\,$169$\,$K and an average surface water ice sublimation rate of ${\dot m_w}$$\,\sim\,$$9\times10^{-7}$$\,$kg$\,$m$^{-2}$$\,$s$^{-1}$ over $R$$\,=\,$2.65$\,-\,$2.75$\,$AU, and $T_{eq}$$\,\sim\,$168$\,$K and ${\dot m_w}$$\,\sim\,$$7\times10^{-7}$$\,$kg$\,$m$^{-2}$$\,$s$^{-1}$ over $R$$\,=\,$2.70$\,-\,$2.80$\,$AU.  The net dust production rates measured in 2010 and 2015 differ by far more than these predicted sublimation rates would indicate, suggesting that other factors besides heliocentric distance are responsible for the observed difference in apparent activity strength (cf.\ Section~4).

\subsection{Comparison to Other Active Asteroids\label{section:other_activeasts}}

\begin{figure*}
	\includegraphics[width=6.5in]{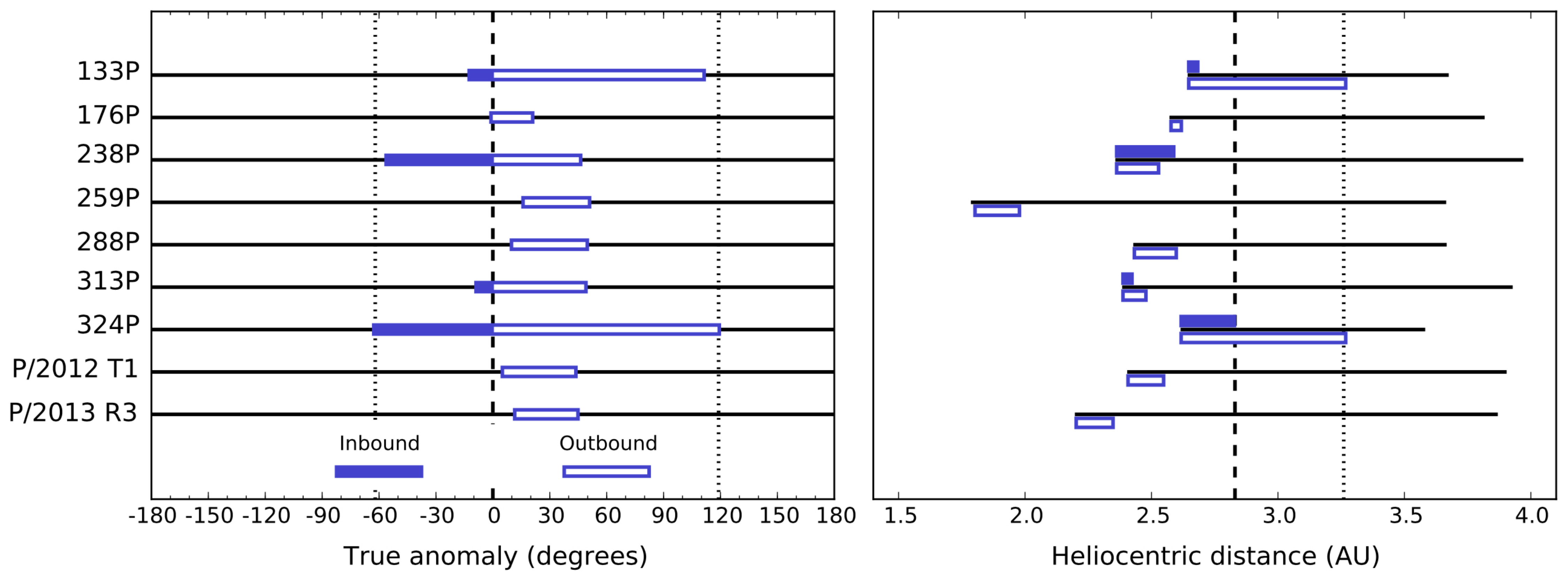}
    \caption{Active ranges \citep[extending between the earliest and latest observations for which activity has been reported; from][and references within]{hsi15a} in terms of true anomaly (left) and heliocentric distance (right) for likely MBCs.  Solid line segments indicate the inbound (pre-perihelion) portion of each object's orbit while outlined segments indicate the outbound (post-perihelion) portion of each object's orbit.  In the left panel, perihelion is marked with a dashed vertical line, while the earliest and latest orbit positions at which activity has been observed for 324P are marked with dotted vertical lines.  In the right panel, horizontal black line segments indicate the heliocentric distance range covered by the orbit of each object, and the most distant positions at which activity has been observed for 324P during the inbound and outbound portions of its orbit are marked with a vertical dashed line and a vertical dotted line, respectively.
}
    \label{figure:fig_obs_all_activeasts}
\end{figure*}

We are interested in comparing 324P's active range, in terms of both orbit position and heliocentric distance, to those of other MBCs.  We plot the confirmed active ranges (where visible dust emission or photometric enhancement has been reported) of all likely MBCs identified to date in Figure~\ref{figure:fig_obs_all_activeasts}.  However, new objects are always discovered while already active, of course, meaning that each object must generally complete at least another full orbit in order for us to observationally constrain the starting point of activity, assuming that it becomes active again and that geometric constraints permit observations over the time period needed to observe the onset of activity.  Successful detections of activity are also dependent on the sensitivity of the telescopes and instruments being used to search for activity.  Therefore, the active regions marked in Figure~\ref{figure:fig_obs_all_activeasts} should be considered lower limits to the full ranges over which activity may be present.

324P now has the largest observed active range of all likely MBCs in true anomaly, with both the earliest and latest observations of activity in terms of orbit position.
It also now has the most distant confirmed inbound activation point ($R$$\,\sim\,$2.8$\,$AU) of all of the MBCs.
Given the large diversity in MBCs in terms of possible activity mechanisms \citep[cf.][]{jew15b} and dynamical properties, however, the significance of these distinctions is unclear at this time.

\section{Discussion}
\label{section:discussion}

To explain the difference in activity strengths measured for 324P from 2010 and 2015 observations, we note that an insulating layer of regolith could cause a lag in the response to temperature changes for an object with subsurface volatile material \citep[e.g.,][]{hsi11}, since it takes time for a Sun-driven thermal wave to propagate through such a layer to the subsurface. This means that 324P's observed 2010 activity (when it was receding from the Sun) could reflect solar heating strength at a smaller effective $R$, while its observed 2015 activity (when it was approaching the Sun) could reflect the solar heating strength at a larger effective $R$.

Mantling could also be responsible for weaker dust production during subsequent perihelion passages \citep[e.g.,][]{ric90}.  \citet{hsi15a} calculated that given an existing rubble mantle thicker than the diurnal thermal skin depth, it would take $\gtrsim$100~years of continuous sublimation to increase the mantle thickness enough to reduce the sublimation rate of water ice by an order of magnitude.  However, an order of magnitude decrease in the sublimation rate could also occur in the transition from freshly exposed ice to ice covered by a rubble mantle with a thickness equal to the diurnal thermal skin depth, a process estimated to require up to $\sim$1~year of ongoing sublimation \citep{hsi15a}.  If 324P's 2010 activity was due to the sublimation of freshly exposed ice, a dramatic decrease in activity strength during the next orbit could be explained by new mantle growth.

Observations of 324P during its current active period that overlap with the true anomaly range of the observations in 2010 are needed to shed light on the possible causes of the lower net dust production rate measured in 2015 relative to 2010.  Numerical dust modeling is also needed to determine the object's true dust production rate at various times (e.g., by taking dust dissipation into account), providing better indications of the sublimation strength at those times.  Numerical dust modeling will also be needed to ascertain 324P's true active range, since the residual dust trail observed in late 2011 likely largely consisted of long-lived large particles, and therefore does not necessarily indicate that ongoing dust production was actually present at the time \citep[e.g.,][]{jew13}.

\section*{Acknowledgements}

This work is based on data gathered with the 6.5 meter Magellan Telescopes at Las Campanas Observatory, as well as with MegaPrime/MegaCam, a joint project of CFHT and CEA/DAPNIA, at CFHT, which is operated by the National Research Council of Canada, the Institute National des Sciences de l'Univers of the Centre National de la Recherche Scientifique of France, and the University of Hawaii.












\label{lastpage}
\end{document}